\documentclass[prl,groupedaddress,twocolumn]{revtex4-2}
\usepackage{bm}
\usepackage{amsmath}
\usepackage{graphicx}
\usepackage{color}
\usepackage{hyperref}

\newcommand{\bracket}[1]{\langle #1 \rangle}

\begin{document}

\title{Thermal Hall Effect of Chiral Spin Fluctuations}

\author{Caitlin Carnahan}
\affiliation{Department of Physics, Carnegie Mellon University, Pittsburgh, Pennsylvania 15213, USA}

\author{Yinhan Zhang}
\affiliation{Department of Physics, Carnegie Mellon University, Pittsburgh, Pennsylvania 15213, USA}

\author{Di Xiao}
\affiliation{Department of Physics, Carnegie Mellon University, Pittsburgh, Pennsylvania 15213, USA}

\date{\today}

\begin{abstract}
Using a two-dimensional square lattice Heisenberg model with a Rashba-type Dzyaloshinskii-Moriya interaction, we demonstrate that chiral spin fluctuations can give rise to a thermal Hall effect in the absence of any static spin texture or momentum space topology.  It is shown by means of Monte Carlo and stochastic spin dynamics simulations that the thermal Hall response is finite at elevated temperature outside of the linear spin wave regime and consistent with the presence of thermal fluctuation-induced nontrivial topology.  Our result suggests that the high-fluctuation phases outside of the conventional regime of magnonics may yet be a promising area of exploration for spin-based electronics.
\end{abstract}

\maketitle

The thermal Hall effect has become an insightful tool for mapping the ground state and low-energy excitations of insulating magnets~\cite{PhysRevLett.104.066403,PhysRevLett.106.197202,PhysRevB.84.184406,Onose297,PhysRevLett.115.106603,Hirschberger106,Watanabe8653}.  In magnetically ordered systems, the energy current is carried by magnons, quantized fluctuations around the magnetic order, and the thermal Hall effect can be understood as the magnon Hall effect.  There are two possible mechanisms for the magnon Hall effect.  If the ground state has lattice translation symmetry, the magnon Hall effect can manifest if the underlying magnon bands possess nontrivial momentum space topology~\cite{zhang2013,mook2014}.  On the other hand, the magnon Hall effect can also appear if the system carries a topologically nontrivial spin texture in real space, in which the magnons can experience an effective magnetic field captured by the scalar spin chirality~\cite{nagaosa2012}
\begin{equation}
\chi_{ijk} = \bm{S}_i \cdot (\bm{S}_{j} \times \bm{S}_{k}) \;.
\end{equation}
This concept has been explored in the topological magnon Hall effect in skyrmion lattices~\cite{hoogdalem2013,mochizuki2014,mook2017}.

While the theory of the magnon Hall effect has been well established, experimentally the thermal Hall effect has been observed to persist at elevated temperatures, even above the ferromagnetic transition temperature~\cite{PhysRevLett.115.106603}.  In this temperature range, the spin dynamics is dominated by large amplitude fluctuations, rendering the magnon description inappropriate.
To date, thermal Hall effect beyond the linear spin wave approximation has remained a poorly understood subject.  Recently, Lee, Han and Lee made the first attempt by formulating the thermal Hall effect based entirely on the spin operators~\cite{lee2015}, therefore bypassing the need to introduce the magnon operator.  However, their actual calculation of the thermal Hall conductivity is still carried out at the mean-field level using either the Holstein-Primakoff or the Schwinger boson formalism, and the role of fluctuations remains unclear.

In this Letter, we employ a set of numerical methods to study the thermal Hall effect beyond the linear spin wave approximation.  We make use of the mediator-agnostic approach introduced in Ref.~\cite{lee2015} and evaluate the thermal Hall conductivity using classical Monte Carlo and stochastic spin dynamics simulations.  Notably, our calculations feature the contribution due to the energy magnetization~\cite{qin2011}, which has not been included in previous numerical studies~\cite{mook2016}. We show that this contribution is crucial to obtaining the correct behavior of the thermal Hall conductivity in the low-$T$ and high field limits.

To investigate the role of spin fluctuations in the thermal Hall effect, we consider a square lattice Heisenberg model with a Rashba-type Dzyaloshinskii-Moriya interaction (DMI).  Within the linear spin wave approximation, the magnon band is topologically trivial in the uniform ferromagnetic phase and the system does not exhibit any magnon Hall effect.  However, we find that a thermal Hall effect can still emerge by increasing thermal fluctuations around the collinear ground state. This fluctuation-driven thermal Hall effect can be connected to the emergent topology due to chiral spin fluctuations, characterized by the \emph{thermodynamic average} of the topological charge $Q = \bracket{\chi_{ijk}}$~\cite{hou2017}. This effect thus shares the same origin as the topological magnon Hall effect~\cite{hoogdalem2013,mochizuki2014,mook2017}, with the crucial difference that it is driven in this case by chiral spin fluctuations as opposed to static spin textures.  Our result reveals the resilience of the thermal Hall effect in the absence of spin ordering and indicates that the high-fluctuation phases outside of the conventional regime of magnonics may yet be a promising area of exploration for spin-based electronics.


\begin{figure}
  \includegraphics[width=0.95\columnwidth]{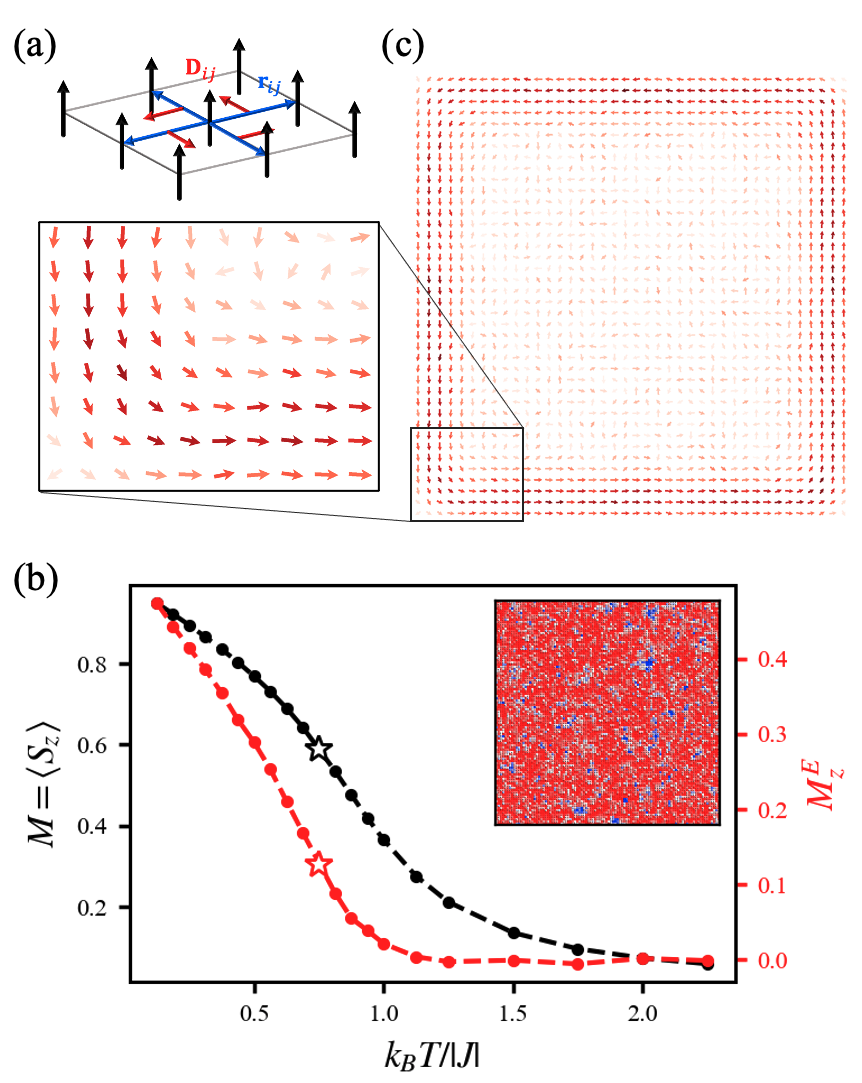}
  \caption{ (a) A simple square lattice model of spin sites. The DM vector (denoted by red arrows) is directed in-plane such that $\bm{D}_{ij} = D \hat{z} \times \hat{r}_{ij}$ where $\bm{r}_{ij}$ (denoted by blue arrows) is the vector describing displacement from site $i$ to nearest-neighbor site $j$. The parameter $D$ describes the strength of the DM interaction. (b) The magnetization $M$ and energy magnetization $M_E$ as a function of $T$ for $D = .3|J|$ and $\bm{B} = .2|J|\hat{z}$. Inset: a snapshot of the local magnetization at $k_B T = .75|J|$ corresponding to the hollow star markers in the plot (red indicates $+\hat{z}$, blue indicates $-\hat{z}$). (c) The spatially-resolved equilibrium average of $\bm{j}^E_i$ for an $N = 40 \times 40$ lattice with $D = .3|J|$, $\bm{B} = .2|J|\hat{z}$,  $k_B T = .75|J|$ and open boundary conditions. The average direction of $\bm{j}^E_i$ is indicated with arrows, while its amplitude is indicated by the arrow color. The time average of the local current
  vanishes in the bulk, but survives on the boundary and circulates in a counter-clockwise fashion in equilibrium. A closer look at the bottom left corner is shown in the inset. }
  \label{fig:square_lattice}
\end{figure}

We begin our discussion by considering a Heisenberg model on a 2D square lattice, described by the Hamiltonian
\begin{equation}\label{eqn:hamiltonian}
H = J\sum_{\left \langle i,j \right \rangle } \bm{S}_{i} \cdot \bm{S}_{j} + \sum_{\left \langle i,j\right \rangle}\bm{D}_{ij}\cdot \left ( \bm{S}_{i}\times \bm{S}_{j} \right ) - \sum_{i}\bm{B} \cdot \bm{S}_{i} \;,
\end{equation}
where the orientation of local magnetic moment at site $i$ is denoted by the dimensionless vector $\bm{S}_{i}$ such that the local magnetization is given by $\mu_B\bm{S}_{i}$ and the interaction parameters have units of energy. The notation $\left \langle i,j \right \rangle$ indicates nearest-neighbor pairs. In this work, we consider $J < 0$ which
leads to ferromagnetic coupling in the exchange term. The Dzyaloshinskii-Moriya vector is directed in-plane and the interaction strength is specified by the parameter $D$, such that $\bm{D}_{ij} = - \bm{D}_{ji} = D\hat{z}\times\hat{r}_{ij}$ where $\bm{r}_{ij}$ is the displacement vector from site $i$ to its nearest neighbor site $j$ as shown in Fig.~\ref{fig:square_lattice}(a).  Despite its simple form, the model described by Eq.~\eqref{eqn:hamiltonian} is known to have a rich phase diagram. At zero temperature, in the presence of a strong external magnetic field, the system exhibits an induced ferromagnetic phase.  As the field strength decreases below $\sim 0.84 D^2/|J|$ the system enters a skyrmion crystal phase, and finally below $\sim 0.27 D^2/|J|$ the system enters a spiral phase~\cite{ezawa2011,banerjee2014,hou2017}.

Let us focus on the collinear ferromagnetic phase.  It has been shown that within the linear spin wave approximation, the DMI cannot enter the magnon Hamiltonian, therefore there can be no magnon Hall effect~\cite{zhang2019}.  Alternatively, since there is only one spin per unit cell, the magnon spectrum consists of only one band, and the momentum space topology must be trivial.  It is clear then that any study of the thermal Hall effect in this model must go beyond the linear spin wave approximation.

The theory of spin-based thermal Hall effect has been developed recently~\cite{lee2015,han2017}, with the advantage of being mediator-agnostic; that is, the presence of a well-defined magnon carrier is not necessary.  In this approach, the central quantity of interest, the energy current $\bm{j}^{E}$, is written as a function of the local magnetic moments by solving the local continuity equation in the energy-conserving limit of zero damping~\cite{lee2015,mook2017}, $dH_{i}/dt + \nabla \cdot \bm{j}_{i}^{E} = 0$, where $H_i$ is the local energy density at site $i$. An expression for the energy current is obtained which, in the present system, can be reduced to $\bm{j}_{i}^{E} = \sum_{j \in nn(i)}\bm{r}_{ij}J^{E}_{ij} + \sum_{j \in nnn(i)}\bm{r}_{ij}J^{E}_{ij}$ where the first sum is taken over nearest neighbors of site $i$, the second sum is taken over the next nearest neighbors of site $i$, $\bm{r}_{ij}$ is the displacement from site $i$ to site $j$, and we've introduced the energy bond current $J^{E}_{ij}$, the form of which is derived in the Supplementary Material~\cite{supp}.

Following the method introduced by Luttinger~\cite{PhysRev.135.A1505}, and recently applied to the spin Hamiltonian in Ref.~\cite{lee2015}, a pseudogravitational potential $\psi$ coupling to the energy density may be introduced in the perturbed Hamiltonian to microscopically approximate the application of a temperature gradient such that $\nabla T = T\nabla \psi$. A crucial observation to be made is that the application of this pseudogravitational potential modifies not only the density matrix, but the energy current definition as well.  Keeping only terms linear in the perturbation, the thermal Hall conductivity is revealed to be a sum of two distinct contributions~\cite{lee2015,han2017},
\begin{equation}
  \label{eqn:thermal_hall}
  \kappa_{xy} = -\frac{\textrm{Tr} \left (\delta \rho\;  j_{x}^{E}\right )}{T \nabla_y \psi} - \frac{\textrm{Tr} \left ( \rho_0  \;\delta j_{x}^{E}\right )}{T\nabla_y \psi} \equiv \kappa_{xy}^{0} + \kappa_{xy}^{1} \;.
\end{equation}
In the above expression, the first term $\kappa_{xy}^{0}$ is given by the well-known Kubo formula,
\begin{equation}
  \label{eqn:kubo}
  \kappa_{xy}^{0} = \frac{N}{k_{B}T^{2}}\int_{0}^{\infty} dt \bracket{j_{x}^{E}(t)j_{y}^{E}(0)} \;,
\end{equation}
where $N$ is the number of spins, and in dropping the subscript $i$ on the energy current, we are taking the spatial average $\bm{j}^{E} = \frac{1}{N}\sum_{i}^{N}\bm{j}_{i}^{E}$. The second term, $\kappa_{xy}^{1}$, which originates from the modification of the energy current, is given by
\begin{equation}
  \label{eqn:correction}
  \kappa_{xy}^{1} = -\frac{2\sum_{i} \bracket{r_{i,y}j_{i,x}^{E}}}{NT} \;.
\end{equation}

Physically, $\kappa_{xy}^1$ represents the circulating component of the energy current that is not observable in the transport experiment and needs to be subtracted~\cite{qin2011}.  This can be seen by noticing that $\sum_i\bracket{r_{i,y}j_{i,x}^{E}}/N$ is essentially half of the energy magnetization $M^E_z = \sum_i\bracket{\bm r_i \times \bm j^E_i}_z/N$~\cite{qin2011}.  In Fig.~\ref{fig:square_lattice}(b) we plot $M^E_z$, along with the spin magnetization, as a function of temperature, obtained via Monte Carlo simulations.  The magnetic field is chosen such that the ground state at $T = 0$ is in the uniform ferromagnetic phase.  In Fig.~\ref{fig:square_lattice}(c) we show a typical profile for the local ensemble-averaged value of the energy current at $k_B T = .75|J|$.  As we can see, the energy current is predominantly limited to a constricted path along the boundary, which leads to a finite energy magnetization.  This boundary energy current was previously discussed by Matsumoto and Murakami using the magnon picture: the confining potential of the boundary exerts a force on the magnon wave packet, which, in the presence of a nonzero Berry curvature, leads to an anomalous velocity along the boundary direction~\cite{PhysRevLett.106.197202,PhysRevB.84.184406}.  Our result here shows that the boundary energy current can still exist even in the absence of any momentum space topology.

The existence of a finite energy magnetization is a strong indication that the system will exhibit a thermal Hall effect. While no net energy current exists in equilibrium, the application of a temperature gradient results in asymmetrical boundary current contributions on either end of the gradient which in turn gives rise to a net thermal Hall current. As we show below, accounting for this contribution is crucial to obtaining the correct behavior of the thermal Hall conductivity in the low-$T$ and high field limits.

To evaluate the thermal Hall conductivity, we perform classical spin simulations.  Numerical evaluation of $\kappa_{xy}^0$ involves integration over the
time-displaced cross-correlation functions of the energy current, and thus requires dynamical information about the system. The dynamics of this model are dictated by the Landau-Lifshitz-Gilbert equation,
\begin{equation}\label{eqn:llg}
  \frac{d\bm{S}_i}{d \tau} = \bm{S}_i \times \frac{\partial H}{\partial \bm{S}_i}  + \alpha  \bm{S}_i \times \Bigl( \bm{S}_i \times \frac{\partial H}{\partial \bm{S}_i}\Bigr) \;,
\end{equation}
presented here in dimensionless form where $\alpha$ is the Gilbert damping parameter and the unitless time $\tau$ yields $d\tau \propto \gamma dt/(1 + \alpha ^2) $ where $\gamma$ is the
gyromagnetic ratio.  To obtain the time-ordered energy currents, we numerically integrate the stochastic version of Eq.~\eqref{eqn:llg} which includes a Gaussian white noise term to establish finite temperature \cite{Skubic_2008} (see Supplemental Material \cite{supp} for details). Evolution of the $N = 80 \times 80$ system is performed with $\alpha = .01$ for $2.5\times 10^{9}$ -- $7.5\times 10^{9}$ time steps, recording the energy currents every $20$ time steps.  To obtain $\kappa_{xy}^1$, we perform single-spin Metropolis Monte Carlo \cite{doi:10.1063/1.1699114} on an $N = 100 \times 160$ lattice with periodic boundary conditions along $x$ and open boundary conditions along $y$. After an equilibration stage of $4\times10^5$ MC sweeps over the lattice, we perform another $6\times10^5$ MC sweeps, recording the local energy current values during each sweep.  We report our results in the normalized form $\tilde{\kappa}_{xy} = \mu_{B}\kappa_{xy}/\gamma k_B$ such that the units of $\tilde{\kappa}_{xy}$ are those of $|J|$.

\begin{figure}
  \includegraphics[width=0.9\columnwidth]{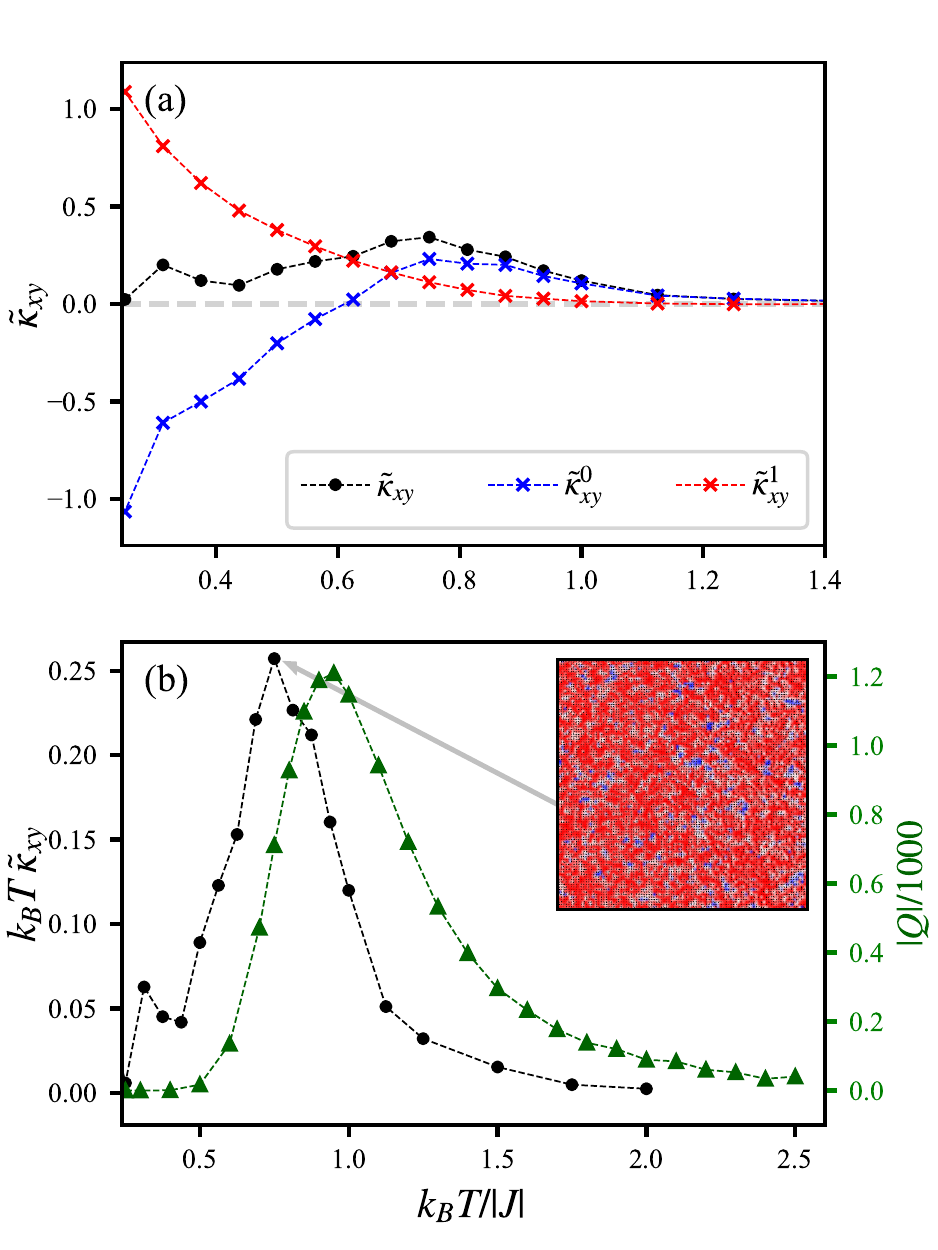}
  \caption{ Temperature dependence of the (a) thermal conductivity $\tilde{\kappa}_{xy}$ (black), Kubo contribution $\tilde{\kappa}_{xy}^{0}$ (blue), energy magnetization correction $\tilde{\kappa}_{xy}^{1}$ (red) and (b) topological charge $|Q|$ per 1000 spins (green triangles) plotted with $k_B T \tilde{\kappa}_{xy}$ at $B/|J| = .2$ with $D = .3|J|$. Inset: a typical snapshot of the local magnetization profile for $B = .2|J|$ and $k_B T = .75|J|$.  }
  \label{fig:temp_fig}
\end{figure}

We first fix $B = .2|J|$ such that the $T \rightarrow 0$ ground state is a collinear ferromagnet. In Fig.~\ref{fig:temp_fig}(a), the thermal Hall conductivity $\tilde{\kappa}_{xy}$ is shown as a function of temperature $T$.  Remarkably, a thermal Hall response --- which is forbidden under the linear spin wave approximation and absent at low $T$ --- can be obtained at elevated temperatures, peaking at approximately $k_B T = .75 |J|$.  The isolated contributions of $\tilde{\kappa}_{xy}^{0}$ and $\tilde{\kappa}_{xy}^{1}$ are included in Fig.~\ref{fig:temp_fig}(a) to illustrate the origins of the peak in $\tilde{\kappa}_{xy}$. As $T \rightarrow 0$ the contributions from each term are opposite in sign and roughly equal in magnitude, leading to suppressed $\tilde{\kappa}_{xy}$ in the low-$T$ region which is consistent with our earlier discussion of the linear spin wave approximation.  We emphasize that this cancellation at low-$T$ can only be obtained through inclusion of the correction due to the energy magnetization; otherwise a finite thermal conductivity would be obtained at $T = 0$.  At approximately $k_B T = .6 |J|$, $\tilde{\kappa}_{xy}^{0}$ switches sign and combines with the decaying $\tilde{\kappa}_{xy}^{1}$ term to produce a peak in $\tilde{\kappa}_{xy}$, before decaying altogether at higher temperatures.

Since the momentum space topology is trivial, the finite $\tilde{\kappa}_{xy}$ could be a response to an underlying static spin texture.  To demonstrate that this is not the case, a snapshot of a typical magnetization profile at the peak in the thermal Hall conductivity is shown in the inset of Fig.~\ref{fig:temp_fig}(b). We see the absence of any periodic spin texture in the lattice and, in fact, the local magnetization is highly fluctuating.

Even though there is no static spin texture, it has been shown that the ensemble average of this highly fluctuating state can still yield a nonzero topological charge $Q$, a quantity defined as the spatial average of $\chi_{ijk}$~\cite{hou2017}.  Hence, on average the magnons can still experience an effective magnetic field~\cite{nagaosa2012} and gives rise to a thermal Hall effect.  To confirm this picture, we calculate the topological charge $Q$ following Ref.~\cite{hou2017} and compare it with the thermal Hall conductivity. The temperature dependence of the thermal Hall signal is most naturally compared to that of the topological charge $Q$ through the quantity $k_B T\tilde{\kappa}_{xy}$, both of which are plotted in Fig. \ref{fig:temp_fig}(b) as a function of $T$.  We see similarity in peak shape and position, indicating that the topological charge in this region is contributing to the thermal Hall signal.
One way to understand this is to consider the thermal fluctuations as giving rise to a transient chiral spin texture, off of which magnons scatter and collectively give rise to a macroscopic response.  We also note that electronic response to fluctuating spin chirality has been recently observed in terms of a topological Hall effect~\cite{wang2019}.

We note that a shift in the peak of the thermal Hall conductivity compared to the topological charge has two sources. First, $\mathcal{O}(S^2)$ terms in the energy current~\cite{supp}, which are amplified by $B$ and thermal fluctuations, do not enter into the topological charge expression. Second, the thermal conductivity is not only a function of the magnitude of $Q$, but also of its time-displaced correlation -- that is, even a strong $Q$ that fluctuates too quickly in time may lead to diminished Hall response.

\begin{figure}
  \includegraphics[width=\columnwidth]{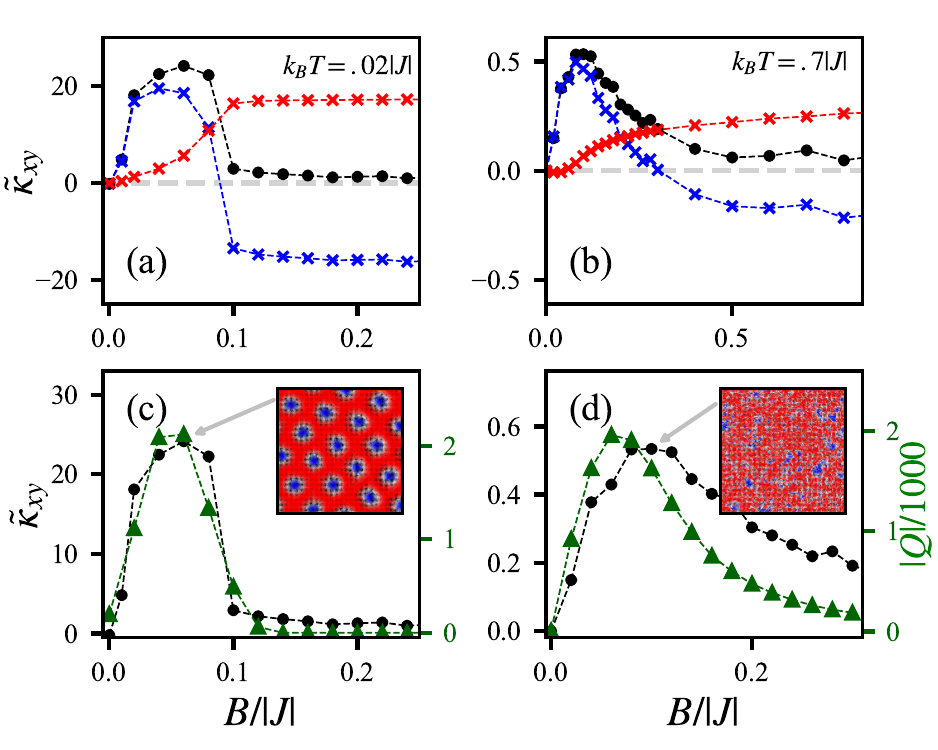}
  \caption{ External field dependence of the thermal conductivity $\tilde{\kappa}_{xy}$ (black), Kubo contribution $\tilde{\kappa}_{xy}^{0}$ (blue), energy magnetization correction $\tilde{\kappa}_{xy}^{1}$ (red) for (a) $k_B T = .02|J|$ and (b) $k_B T = .7|J|$. Thermal conductivity $\tilde{\kappa}_{xy}$ (black) is again plotted with topological charge $|Q|$ per 1000 spins (green triangles) for (c) $k_B T = .02|J|$ and (d)$k_B T = .7|J|$. Insets: a typical snapshot of the local magnetization profile for (c) $B = .06|J|$ and $k_B T = .02|J|$  (d) $B = .1|J|$ and $k_B T = .7|J|$.  The DMI strength is fixed to $D = .3|J|$ throughout.}
  \label{fig:b_fig}
\end{figure}

We may also control the spin fluctuations in our system by fixing $T$ and varying the external field strength.  In the limit of $B \to \infty$ for any finite $T$, the magnetization should saturate and fluctuations are entirely suppressed. We first look at the low-$T$ situation by fixing $k_B T = .02|J|$.  The thermal conductivity is shown in Fig.~\ref{fig:b_fig}(a) and plotted against $Q$ in Fig.~\ref{fig:b_fig}(c). As mentioned previously, the system undergoes several phase transitions by increasing $B$ near $T = 0$. The spiral phase obtained by setting $B = 0$ transforms into a skyrmion crystal phase at  $B\approx .025|J|$; in this phase, both $\tilde{\kappa}_{xy}$ and $Q$ reach their maximum value and eventually vanish with increasing $B$ at $B \approx .1|J|$, just beyond the onset of the induced ferromagnetic phase. At such a low $T$, thermal fluctuations are minimal in the collinear phase and the thermal conductivity vanishes in accordance with the linear spin wave result. There is stunning agreement between the thermal conductivity and topological charge profile at this temperature, highlighting the necessity of the energy magnetization correction to calculate the thermal response accurately. In the skyrmion crystal phase, we see that the peak in the conductivity is due to collaborative contributions by both the bulk and boundary current response. However, at the onset of the induced ferromagnetic phase, the Kubo term switches sign abruptly and we obtain asymptotic cancellation of the terms, and therefore vanishing conductivity, for large values of $B$.

The story is quite similar for the field dependence in the elevated temperature case $k_B T = .7|J|$, shown in Fig.~\ref{fig:b_fig}(b) and (d). A finite $B$ is necessary to break the time-reversal symmetry in the system and produce a thermal Hall response, but at higher $B$ values the system approaches an induced ferromagnetic phase in which $\tilde{\kappa}_{xy}$ must vanish. The thermal conductivity coincides with the presence of topological charge in the system as seen in Fig. \ref{fig:b_fig}(b). However, the peak in the thermal conductivity occurs in the absence of any periodic spin texture (see the inset of Fig.~\ref{fig:b_fig}(d)), as opposed to the low-$T$ peak which appears in the skyrmion crystal phase (see inset of Fig. \ref{fig:b_fig}(c)). This result indicates that thermal Hall response due to magnons scattering off of the underlying topological spin texture can be observed in the case of a dynamic spin texture due to chiral fluctuations, as well as a static spin texture.

In summary, we have extended existing numerical methods to demonstrate the significant role of the energy magnetization in shaping the behavior of thermal response in insulating chiral magnets. In particular, our classical simulations have shown that a definite thermal Hall signal can persist alongside a finite topological charge even in the absence of a robust magnon carrier. This result indicates that as magnon coherence is degraded by thermal fluctuations, paramagnons that remain can support a thermal Hall signal through collective response to a dynamic topological spin texture. This finding suggests that, for spin-based electronics, there is fertile ground for transport in the high-fluctuation phases outside of the linear spin wave regime that hosts a long-range magnon carrier.

We wish to thank Claudia Mewes, Alexander Mook, Satoshi Okamoto, Jiadong Zang, and Jian-Gang Zhu for helpful discussion. This work is supported by the Defense Advanced Research Project Agency (DARPA) program on Topological Excitations in Electronics (TEE) under grant number D18AP00011 and by AFOSR MURI 2D MAGIC (FA9550-19-1-0390).  This work used the Extreme Science and Engineering Discovery Environment
\cite{xsede},
which is supported by National Science Foundation grant number ACI-1548562. Specifically, this work made use of the Bridges and Bridges-2 resources at the Pittsburgh Supercomputing Center through allocation No. TG-PHY190042.

\textit{Note added}.---A recent work based on the Schwinger boson formalism has also found a thermal Hall effect beyond the linear spin wave approximation for the spin Hamiltonian~\eqref{eqn:hamiltonian}, and its origin has been attributed to magnon-magnon interactions~\cite{PhysRevB.102.214421}.

\bibliography{aps_square_b}

\end{document}